\documentclass{emulateapj}

\slugcomment{{\sc Accepted to ApJ:} August 4, 2010} 
\usepackage{graphicx}
\usepackage{amsmath}

\usepackage{epstopdf}
\usepackage{natbib}

\begin{document}

\title{Six High-Precision Transits of OGLE-TR-113b\altaffilmark{1}}

\author{E. R. Adams\altaffilmark{2}, M. L\'opez-Morales\altaffilmark{3}, J. L. Elliot\altaffilmark{2,4},  S. Seager\altaffilmark{2,4},  D. J. Osip\altaffilmark{5}}

\altaffiltext{1}{This paper includes data gathered with the 6.5 meter Magellan Telescopes located at Las Campanas Observatory, Chile.}
\altaffiltext{2}{Department of Earth, Atmospheric, and Planetary Sciences, Massachusetts Institute of Technology, 77 Massachusetts Ave., Cambridge, MA, 02139}
\altaffiltext{3}{Department of Terrestrial Magnetism, Carnegie Institution of Washington, 5241 Broad Branch Road NW, Washington, DC 20015-1305; Hubble Fellow}
\altaffiltext{4}{Department of Physics, Massachusetts Institute of Technology, 77 Massachusetts Ave., Cambridge, MA, 02139}
\altaffiltext{5}{Las Campanas Observatory, Carnegie Observatories, Casilla 601, La Serena, Chile}
\begin{abstract}

We present six new transits of the hot Jupiter OGLE-TR-113b observed with MagIC on the Magellan Telescopes between January 2007 and May 2009. We update the system parameters and revise the planetary radius to $R_p=1.084\pm0.029~R_J$, where the error is dominated by stellar radius uncertainties. The new transit midtimes reveal no transit timing variations from a constant ephemeris of greater than 13$\pm$28 seconds over two years, placing an upper limit of $1$-$2~M_{\oplus}$ on the mass of any perturber in a 1:2 or 2:1 mean-motion resonance with OGLE-TR-113b. Combining the new transit epochs with five epochs published between 2002 and 2006, we find hints that the orbital period of the planet may not be constant, with the best fit indicating a decrease of $\dot{P}=-60\pm15$ milliseconds per year. If real, this change in period could result from either a long-period (more than 8 years) timing variation due to a massive external perturber, or more intriguingly from the orbital decay of the planet. The detection of a changing period is still tentative and requires additional observations, but if confirmed it would enable direct tests of tidal stability and dynamical models of close-in planets.

\end{abstract}


\section{Introduction}

In the decade since detecting the first transiting exoplanet \citep{Charbonneau2000, Henry2000}, over ninety transiting systems have been identified. As instrumentation, observations and analysis techniques have improved, so has the precision with which parameters of transiting systems can be measured, prompting theoretical work to extract more information from each light curve. One focus is on how light curves are modified by gravitational interactions with other planets or tidal interactions with the star. For example, variations in the inclination or duration of successive transits would indicate a precessing planetary orbit, potentially from a second planet \citep{MiraldaEscude2002}. Periodic midtime variations could indicate additional planets or moons \citep{Holman2005, Agol2005, Heyl2007, Ford2007, Simon2007, Kipping2009a, Kipping2009b}. A long-term decrease in the orbital period of the transiting planet could indicate orbital decay, a consequence of tidal dissipation \citep{Sasselov2003,Patzold2004,Carone2007,Levrard2009}. 

In this work we present six new transit light curves of the hot Jupiter OGLE-TR-113b, with a focus on the timing effects. In Section~2 we describe the observations, data analysis and light curve fitting. Section~3 describes our timing analysis and the physical implications of our findings. We discuss our results in Section~4.

\section{Observations}
\label{obs}

OGLE-TR-113b was reported by \citet{Udalski2002c} as a planet candidate transiting a K-dwarf [I=14.4; RA(J2000)=10:52:24.40, Dec(J2000)=--61:26:48.5]. Its planetary nature was confirmed by \citet{Bouchy2004} and \citet{Konacki2004}. OGLE-TR-113 is an excellent target for differential photometry, since it is located in a dense region of the sky toward the Galactic plane and has numerous nearby, bright comparison stars. We observed six transits of OGLE-TR-113b as part of a larger campaign to detect transit timing variations of OGLE planets \citep{Adams2010PhD}. All transits were observed in the Sloan $i'$--band with the dual-CCD instrument MagIC on the Magellan Telescopes, located at Las Campanas Observatory in Chile. 

Both MagIC CCDs have low readout noise (about 6 e- per pixel), small fields-of-view and high-resolution pixels (SITe: $142\arcsec\times142\arcsec$, or $0.\arcsec 069$ per pixel; e2v: $38\arcsec\times38\arcsec$, or $0.\arcsec037$ per pixel), which both minimizes blends and produces stellar images that are spread over many pixels (typical FWHM=10--20 pixels depending on seeing and binning). This last feature reduces differential pixel response effects and increases the total number of photons that can be collected per frame, without requiring defocusing. The SITe gain was 2.0 e-/ADU, while the e2v gain was 2.4 e-/ADU in 2008 and 0.5 e-/ADU in 2009 (due to engineering changes). The main advantage of the e2v is its frame-transfer capability: the readout time per frame is only 5s in standard readout mode and 0.003s in frame transfer mode, surpassing the 23s readout time of the SITe chip and other conventional CCDs.

We used MagIC-SITe for two transits in January 2007 and February 2008, henceforth denoted by their UT dates as 20070130 and 20080225. MagIC-e2v was used for four additional transits between April 2008 and May 2009 (20080424, 20080514, 20090315, and 20090510). The nightly sky conditions ranged from photometric to partly cloudy. Therefore, exposure times were adjusted from 10-120 seconds to maintain at least $10^6$ photons per frame for both the target and one or more comparison stars. All transits were observed at air mass $<$ 1.7 to minimize differential color effects. The positions of the stars remained within 5-10 pixels each night, while seeing values oscillated from 0.\arcsec4 to 0\arcsec.7. All transits were observed with $1\times1$ binning, except for 20090510, which was binned $2\times2$. Observations lasted from 3.5 to 5 hours to include the full transit and out-of-transit baseline. Observations were cut short for 20070130 due to clouds. We also truncated the 20080424 light curve due to a strong systematic slope, which we were unable to remove (see Section~\ref{section:photometry}). 

Given the utmost importance of accurate timing, we took special care to correctly record the time in the image headers. The MagIC-SITe and 2008 MagIC-e2v times came from the network server, which we verified to be synchronized with the observatory's GPS clocks at the beginning of each night. The 2009 MagIC-e2v times came from an embedded PC104 computer, which receives unlabeled GPS pulses every second, and is synchronized to the observatory's GPS every night. The intrinsic error for all header times is thus much less than a second.

\subsection{Photometry}
\label{section:photometry}

All data were overscan-corrected and flatfielded using standard IRAF routines\footnote{
IRAF is distributed by the National Optical Astronomy Observatory, operated by the Association of Universities for Research in Astronomy, Inc., under cooperative agreement with the National Science Foundation}. Simple aperture photometry, using the IRAF ${\it apphot}$ package, provided optimal results, since the target and comparison stars appear well isolated on all frames. We ran a wide grid of apertures and sky annuli to locate the optimal values, which minimized the out-of-transit photometric dispersion of the differential light curves. The comparison stars were iteratively selected to be similar in brightness to the target, photometrically stable and un-blended. We examined 10-20 stars for each transit, using 1, 3, or 10 in the final light curves, depending on the night. The best aperture radii were 4-24 pixels, while the sky annuli had inner radii of 30-120 pixels and 10-pixel widths. A few images were discarded because of low counts or saturation. For 20080424, a 4-pixel aperture (significantly smaller than the 15-pixel FWHM radius) was adopted to eliminate a slope in the light curve before and during transit, but failed to correct the slope after transit.

We examined the out-of-transit baseline of each light curve for systematic trends with respect to air mass, seeing, telescope azimuth, ($x$,$y$) pixel location, and time \citep[for more details see][]{Adams2010a}. No trends were found in three light curves (20080225, 20080514, and 20090510), and a fourth (20080424) was not detrended, but rather truncated after transit since the slope could not be modeled against any variable. We removed a slope in telescope azimuth (0.022$\%$ deg$^{-1}$) from the 20070130 light curve; the transit midtime was unchanged but the photometric scatter improved and the depth now agrees with the other light curves. We also removed a slope in seeing from the 20090315 light curve ($-0.16\%$ pixel$^{-1}$), observed with variable sky transparency conditions. 

\begin{figure}
\includegraphics*[scale=0.35]{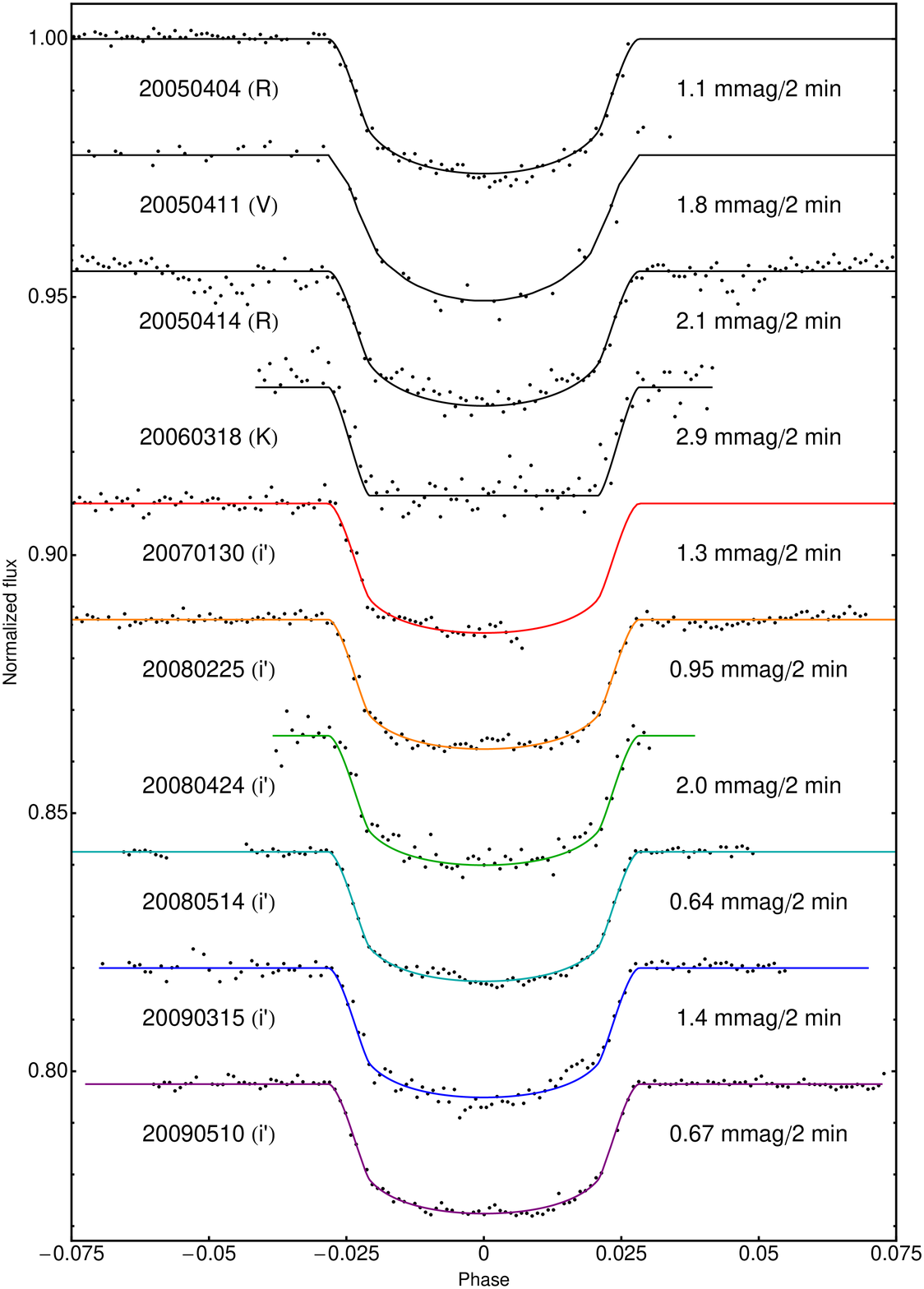}
\caption{Ten transits of OGLE-TR-113b. All transits, labeled by date and filter, are binned to 2 minutes and plotted against orbital phase. The top four transits are from \citet{Gillon2006}, \citet{Pietrukowicz2010}, and \citet{Snellen2007}; the rest are new. Solid lines show the best joint-fit model, and we report the standard deviation of the residuals.}
\label{fig:ogle113transits}
\end{figure}

The photometric data for each new transit is excerpted in Table~\ref{table:ogle113data}. All light curves are illustrated in Figure~\ref{fig:ogle113transits}, binned to 2 minutes for visual comparison, though the full data were used in all fits. 

Our analysis also includes four literature light curves: 20050404 and 20050414 in $R$-band \citep{Gillon2006}, 20050411 in $V$-band \citep{Pietrukowicz2010, Diaz2007}, and 20060318 in $K$-band \citep{Snellen2007}. The first three light curves were kindly provided by the lead author of each paper. The original K-band data were lost to a hard-drive crash (Snellen; private communication), so we reconstructed the unbinned light curve from Figure 3 of their paper. We were able to extract the points within a few seconds and reconverted the phased data to UTC times with zero phase at 2006-03-18 04:56:00 UTC (transit epoch 1038 from the quoted ephemeris). Finally, we did not re-fit the OGLE survey photometry, but instead used the transit epoch from \citet{Konacki2004} in our timing analysis (see Section~\ref{section:ogle113timing}).

\subsection{Light curve fitting}
\label{section:ogle113fitting}

All light curves in Figure~\ref{fig:ogle113transits} were jointly fit using the \citet{Mandel2002} algorithm, with the MCMC method of \citet{Carter2009}, as described in \citet{Adams2010a}. We assumed quadratic limb-darkening, with initial values from \citet{Claret2000,Claret2004}, using $T=4804~K$, $\log{g}=4.52$, $[M/H]=0$, and $v_{micro}=2$~km s$^{-1}$ \citep{Santos2006}. We assumed zero limb-darkening for the flat-bottomed $K$-band transit, following \citet{Snellen2007} but in contradiction with \citet{Claret2000}\footnote{The limb-darkening coefficients used are: $u_{1,R}=0.52\pm0.02$,  $u_{2,R}=0.19$ (fixed), $u_{1,V}=0.77\pm0.04$, $u_{2,V}=0.10$ (fixed), $u_{1,i'}=0.40\pm0.02$, $u_{2,i'}=0.23$ (fixed), $u_{1,K}=0$ (fixed), $u_{2,K}=0$ (fixed).}. Throughout we adopted $M_*=0.78\pm0.02~M_{\odot}$, $R_*=0.77\pm0.02~R_{\odot}$ and $M_{p}=1.32\pm0.19~M_J$ \citep{Santos2006}, and zero obliquity, oblateness and orbital eccentricity for the planet. The orbital period was fixed to $P=1.43248$ days after tests showed little effect on the transit fit results; we similarly fixed $u_2$ for each filter. 

The best model fit and corresponding parameter distributions were derived from three independent Markov chains of $10^6$ links each, discarding the first 50,000 links. We fit values for the planet/star radius ratio, $k=0.1447^{+0.0006}_{-0.0005}$, semimajor axis in stellar radii, $a/R_*=6.47^{+0.06}_{-0.09}$, and orbital inclination, $i=89.0^{+1.1}_{-0.7}$. The fitted transit midtimes are shown in Table~\ref{table:ogle113ominusc}. We derive values for  the impact parameter, $b=0.11^{+0.07}_{-0.09}$, total transit duration $T_{14}=9647^{+31}_{-27}$ s, planetary radius $R_p=1.084\pm0.029$ $R_J$ ($R_J=71,492$ km), and semimajor axis, $a=0.02315^{+0.00064}_{-0.00067}$ AU. The errors are discussed in Section~\ref{section:ogle113systematics}; note that errors on $R_p$ and $a/R_*$ incorporate stellar radius uncertainties.

Additionally, each transit was independently fit to check for parameter variations over time. Over eight years, all values for $k$, $i$, $a/R_*$, and $T_{14}$ are consistent within 1-$\sigma$ of each other and of the joint-fit parameters.

\subsection{Systematic errors}
\label{section:ogle113systematics}

Light curve systematics caused by atmospheric/instrumental effects, stellar noise, planetary satellites or star spots can affect the correct determination of transit parameters. Correlated noise often introduces residuals 2-3 times larger than Gaussian noise. Such effects have been discussed in detail by \citet{Pont2006} and more recently by \citet{Carter2009} in the specific context of transit midtime determination.

We evaluated how systematics increase the errors estimated from MCMC parameter distributions using two methods: time-averaged residuals and residual permutation \citep{Pont2006, Southworth2008}, as detailed in \citet{Adams2010a}. These methods give similar results, with errors up to twice those of the MCMC fits, depending on the light curve. The errors reported in Table 2 and Section~\ref{section:ogle113systematics} correspond to the 68.3$\%$ credible interval limits from the MCMC distributions, scaled upwards by the excess noise factors for each parameter from the residual permutation method. Note that these errors may prove in fact to be too conservative in the analysis of transit timing variations in Section~\ref{section:ogle113timing}.

Finally, \citet{Snellen2007} noted that OGLE-TR-113 shows low-amplitude, long-period photometric variability, so we have examined whether stellar variability could account for correlated features in our light curves. For example, the bump before mid-transit in the 20080514 light curve (Figure~\ref{fig:ogle113transits}) could be interpreted as a star spot, but we discard that hypothesis after confirming that it coincides with a temporary rapid increase in seeing.

\section{Timing}
\label{section:ogle113timing}

To ensure uniform timing analysis, all times in Table~\ref{table:ogle113ominusc} are reported in Barycentric Julian Days, using the UTC-TT and TT-TDB conversions ($BJD_{TDB}$) after \citet{Eastman2010}\footnote{UTC, or Coordinated Universal Time, is typically recorded in the image headers. TT, or Terrestrial Time, is TT=UTC+32.184s+$\Delta$T, where $\Delta$T is the number of leap seconds since 1972. TDB, or Barycentric Dynamical Time, includes millisecond-level relativistic effects.}. The \citet{Gillon2006} light curves were supplied with UTC times, which we converted to $BJD_{TDB}$ along with our transits. We similarly converted the reconstructed UTC times from the \citet{Snellen2007} light curve. We added 64.184s to the data from \citet{Pietrukowicz2010} and \citet{Konacki2004}, after confirming that the UTC-TT conversion had not been applied.

Four attempts to fit a transit midtime ephemeris are illustrated in Figure 2. Panel A shows the \citet{Gillon2006} ephemeris, derived using three epochs: 20020220 (OGLE survey), 20050404 and 20050414 \citep{Gillon2006}. After accounting for the 64.184s UTC-TT conversion, the ephemeris is
\begin{align}
T_{C}(N)=2453464.61740(10) [BJD_{TDB}] \nonumber \\
+1.43247570(130)N,
\label{ogle113eqn1}
\end{align}
where $T_C$ is the predicted transit midtime in $BJD_{TDB}$, the first term is the reference midtime, $T_0$, and the second term is orbital period, $P$, times $N$, the number of elapsed transits since $T_0$. Although the original data is well fit, this ephemeris cannot account for the full data set, giving a reduced $\chi^2=26$.

Panel B shows a new ephemeris fit to all eleven epochs:
\begin{align}
T_{C}(N)=2453464.61762(27) [BJD_{TDB}] \nonumber \\
+1.43247425(34)N,
\label{ogle113eqn2}
\end{align}
with a reduced $\chi^2=1.6$. (The errors in Equation 2 have been re-scaled by $\sqrt{1.6}=1.3$ to account for
the deviation of the fit from $\chi^2=1$.) Although this equation produces a fair fit to the data, it fails to reproduce some of the most accurate transit midtimes, e.g. the 20090510 transit, which deviates by more than 2-$\sigma$ from the fit.

Panel C shows a third ephemeris fit to only the 2007-2009 epochs: 
\begin{align}
T_{C}(N)=2453464.61857(15) [BJD_{TDB}] \nonumber \\
+1.43247315(13)N.
\label{ogle113eqn3}
\end{align}
This fit has a reduced $\chi^2=0.3$ and low errors, but cannot account for the 2002-2006 epochs. 

The orbital period in Equation 3 differs by $-0.22\pm0.11$s from the period in Equation 1, with most of the error associated with the sparser, lower-precision 2002-2005 data. Possible explanations for this period discrepancy are: (1) the timing errors of the new transits have been underestimated --- however, the very low reduced $\chi^2$ of Equation 3 suggests the opposite; (2) the literature midtimes contain yet-unaccounted-for systematic errors, a possibility which should be investigated, since timing errors of hundreds of seconds have been reported before \citep{Adams2010a, Johnson2010}; or (3) the observed orbital period of OGLE-TR-113b has decreased with time. The third explanation, which must be tested with additional data, could be due to a yet-undetected binary star companion \citep{Montalto2010} with modulation period longer than the eight years of observations, or to a real, linear decrease in the period due to orbital decay of the planet. 

If the change in period is real, it is also rapid. Assuming for simplicity a linear rate of change, all 11 transit epochs are well fit if the orbital period decreases by $\dot{P}= -60\pm15$ ms yr$^{-1}$, with
%\begin{equation}
\begin{align}
T_{C}(N)=2453464.61873(8) [BJD_{TDB}] \nonumber \\
 +1.43247426(11)N+\delta P*N(N-1)/2,
\label{ogle113eqn4}
%\end{equation}
\end{align}
where $\delta P=-(2.74\pm0.66)\times10^{-9}$ days is the amount by which the orbital period changes per orbit, assuming $P(N)=P_0+\delta P\times N$; note $\dot{P}=\delta P\times365.25/ P_0$. The reduced $\chi^2=0.6$ indicates this is a better fit to all epochs by a factor of 2.7 than the constant-period fit in Equation 2. Only the 20050411 epoch deviates by more than 1-$\sigma$ (1.7), with residuals of $-80\pm47$s. 

As another check on the robustness of this detection, we also fit for the average number of periods between each pair of midtimes. We find a similar rate, $\dot{P}=-76\pm11$ ms yr$^{-1}$, with reduced $\chi^2=0.7$; the fit assuming a constant average period has reduced $\chi^2=2$.

\subsection{Theoretical orbital decay}

We now briefly explore the theoretical implications of assuming the observed period decay is real and due to orbital decay. Theoretical studies of the closest-in planets have until recently focused on OGLE-TR-56b, for many years the shortest-period exoplanet known. The estimated timescales of orbital decay vary greatly, due to different model assumptions and to uncertainties in poorly-constrained parameters. In particular, estimates of the stellar tidal dissipation factor, $Q_*$, span five orders of magnitude \citep{Patzold2004}. Models that rely on dissipation by turbulent viscosity find lifetimes of a few billion years. \citet{Sasselov2003} estimates a lifetime for OGLE-TR-56b of 0.77 Gyr, implying the orbital period changes by $\dot{P}= 2$~ms yr$^{-1}$ (0.1--5 ms yr$^{-1}$ under different model choices). \citet{Patzold2004} and \citet{Carone2007} find a longer lifetime for OGLE-TR-56b, 3-3.5 Gyr. A recent study by \citet{Levrard2009} challenges the assumption that stable or unstable tidal equilibrium modes are applicable, and finds fast orbital decay rates for most transiting planets. Their remaining lifetime for OGLE-TR-56b is only 7 Myr, assuming $Q_*=10^6$ (0.7 Myr to 70 Gyr for $Q_*=10^5$-$10^{10}$).

Adopting the \citet{Levrard2009} prediction (their equation 5), the lifetime for OGLE-TR-113b is 98 Myr, assuming $Q_*=10^6$ (9.8 Myr to 980 Gyr for $Q_*=10^5$-$10^{10}$). Assuming a constant decay rate and that the stellar Roche lobe radius is reached after the period has decreased by 0.98 days (to $P_{Roche}=0.45$ d), the fastest predicted linear orbital decay rate is 9.8 ms yr$^{-1}$, if $Q_*=10^5$. Our tentative period change is six times faster, and could be reproduced by the above model with $Q_*=16000^{+5000}_{-3000}$, six times smaller than the theoretical lower estimate. 

\subsection{Mass limits on short-period perturbers}

To place upper mass limits on additional objects near OGLE-TR-113b, we numerically integrated objects with periods from 0.3 to 6.3 days, using the approach of \citet{Steffen2005} as described in \citet{Adams2010a}. We restricted the analysis to the 6 new transits, which can acceptably be fit with a constant period, an underlying assumption of the method. In Figure 3 we show results for perturbers on both initially circular ($e=0$) and slightly eccentric ($e=0.05$) orbits, with similar constraints. The strongest constraints are placed near the 1:2 and 2:1 mean-motion resonances, where objects as small as 1-2 $M_{\oplus}$ would have been detectable. Several additional resonances, including the 5:2 and 3:1, also exclude objects larger than 2-20 $M_{\oplus}$. We therefore find no evidence for timing variations caused by close companion planets.

\section{Discussion}
\label{section:ogle113conclusions}

We have measured 6 new transits of OGLE-TR-113b, which provide the highest-quality data for this planet to date. The new transits have timing precisions from 9-28s and photometric precisions as good as 0.6-0.7 millimagnitudes in 2 minutes (20080514 and 20090510). The error on the refined planetary radius, $R_p=1.084\pm0.029~R_J$, is dominated by stellar radius uncertainties.

Observations of OGLE-TR-113b span eight years of data (2002-2009) after including five previously published epochs. We find no evidence of short-period timing variations, and therefore no sign of other planets in close proximity to OGLE-TR-113b. However, there are hints that the orbital period calculated from the 2007-2009 epochs does not agree with orbital period calculated from earlier epochs. One explanation for this discrepancy is that the period is decreasing by $\dot{P}=-60\pm15$ ms yr$^{-1}$.

Given the small number of data points, this detection is still quite tentative, and systematic effects must be carefully considered before any claim is secure. However, the fact that the transit parameters are all self-consistent, as noted in Section~\ref{section:ogle113fitting}, suggests that systematics do not play a large role. In particular, we find no variation greater than 1-$\sigma$ for any of the non-timing parameters when each light curve is independently fit. In addition, the fits in Equations 2 and 4 have reduced $\chi^2$ values significantly smaller than 1.0, indicating that we may be overestimating, not underestimating, the timing errors in Table 2.

If the detection is real, one possibility is a wide-period, possibly stellar-mass companion producing the perturbations over decades-long time scales, as recently noted by \citet{Montalto2010}. Another possibility is that the orbital period of the planet is decaying. Assuming a linear decay rate, the observed rate of period change agrees with the theoretical estimates of planetary lifetime of \citet{Levrard2009} if the host star has a tidal energy dissipation factor $Q_*=16000^{+5000}_{-3000}$.

This preliminary hint of orbital period decay needs to be verified with observations of the OGLE-TR-113 system, which next becomes observable in January 2011. If confirmed, this system has the potential of becoming a key laboratory to improve current dynamical and tidal stability models of close-in planetary systems, and to provide observational constraints on $Q_*$.

\acknowledgements

E.R.A. received support from NASA Origins grant NNX07AN63G. M.L.M. acknowledges support from NASA through Hubble Fellowship grant HF-01210.01-A/HF-51233.01 awarded by the STScI, which is operated by the AURA, Inc. for NASA, under contract NAS5-26555. We thank Adam Burgasser, Josh Carter, Dave Charbonneau, Dan Fabrycky, Jenny Meyer, Paul Schechter, Brian Taylor, Josh Winn, and the Magellan staff for contributions to the observations and analysis; Michael Gillon, Pawel Pietrukowicz, Ignas Snellen and Andrzej Udalski for helpful communications regarding their data; and an anonymous referee for improving this manuscript.

\clearpage

%---------------------------------------------------------------------------
\bibliographystyle{apj}
%\bibliography{main}

%---------------------------------------------------------------------------

\newpage

\clearpage

\begin{figure}
\includegraphics*[scale=0.4]{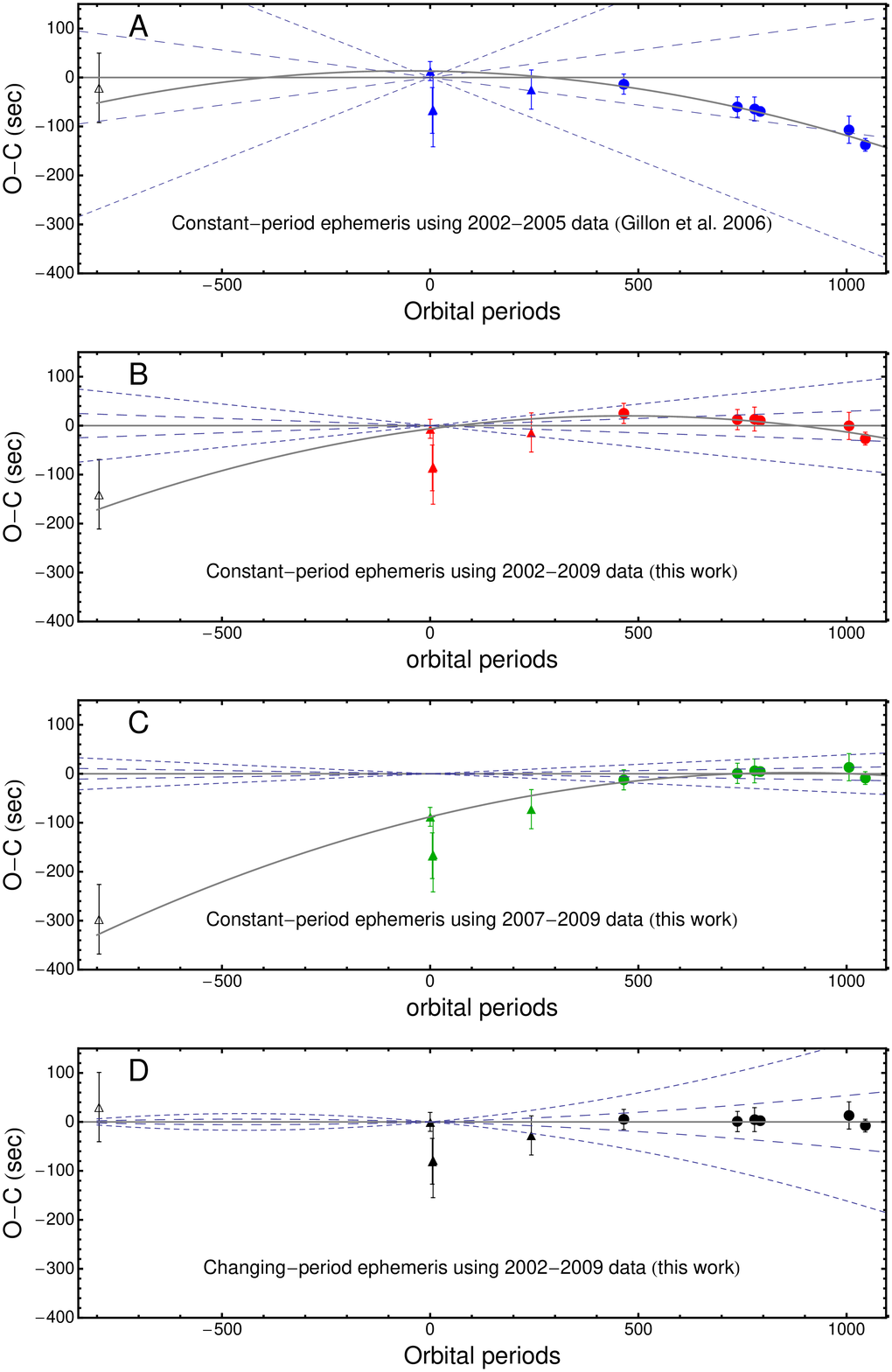}
\caption{Observed minus calculated midtimes for OGLE-TR-113b. Literature times are plotted as triangles and new data as circles; the jointly-fit transits are filled symbols. Each panel shows a different ephemeris: (A) constant period from \citet{Gillon2006}, (B) constant period using all data, (C) constant period using only 2007-2009 data, (D) linearly-changing period using all data.  The curving gray line in (A)-(C) shows the changing-period ephemeris from (D). In all panels dashed lines show 1-$\sigma$ and 3-$\sigma$ errors on the period; those in (D) also incorporate the error on $\dot{P}$. }
\label{fig:ogle113ominusc}
\end{figure}

\begin{figure}
\includegraphics*[scale=0.7]{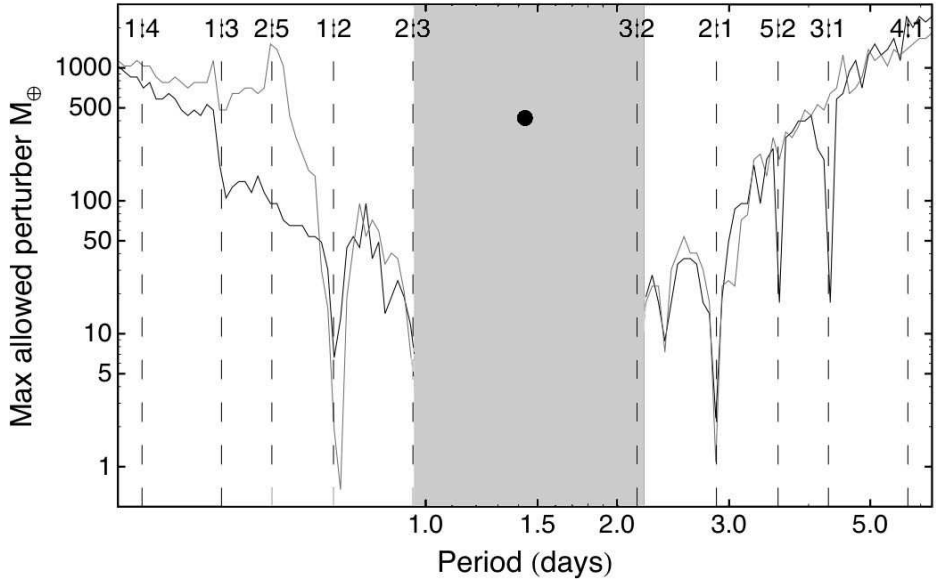}
\caption{Upper mass limit on additional planets with initial eccentricities $e_c=0.05$ (black) and $e_c=0.0$ (gray). The central point shows OGLE-TR-113b. Near the internal 1:2 mean-motion resonance, objects as small as 1-2~$M_{\oplus}$ would have been detectable. The shaded grey region shows the instability region for a $1~M_{\oplus}$ companion, following \citet{Barnes2006}.}
\label{fig:ogle113maxmass}
\end{figure}

\clearpage

%%%%%%%%%%%%%%% Tables

\begin{deluxetable}{l l l l }
\tablewidth{0pt}
\tabletypesize{\scriptsize}
\tablecaption{Transit Flux Values\tablenotemark{a}}
\tablehead{ Mid-exposure (UTC)	& Mid-exposure ($BJD_{TDB}$)	& \textrm{Flux}  		& Error}
\startdata
2454130.599893	&	2454130.601328	&	0.9989367	&	0.001434 \\
2454130.60128	&	2454130.602715	&	1.002827		&		0.001434 \\
2454130.602397	&	2454130.603832	&	1.003337		&	0.001434 \\
2454130.603408	&	2454130.604843	&	1.000978		&	0.001434 \\
2454130.604375	&	2454130.60581	&	0.998505		&	0.001434 \\
\nodata
\enddata     
\tablenotetext{a} {This table is available in its entirety in a machine-readable form in the
online journal. A portion is shown here for guidance regarding its form and content.}
\label{table:ogle113data}
\end{deluxetable}

\begin{deluxetable}{l c c c c c c }
\tablewidth{0pt}
\tabletypesize{\scriptsize}
\tablecaption{Transit Midtimes and Residuals }
\tablehead{Transit 	&Midtime ($BJD_{TDB}$)		& Number & O-C (s)\tablenotemark{a}&  $\sigma$	& O-C (s)\tablenotemark{b}&  $\sigma$}
\startdata
20020220		& ($2452325.79897\pm0.00082$)\tablenotemark{c}		& -795 	& $-140\pm71$ 	& -2.0		& $30\pm 71$ 		& 0.4\\
20050404		& $2453464.61755\pm0.00022$  					& 0 		& $6\pm19 $		& -0.3		& $0\pm19$ 		& 0.0 \\
20050411		& $2453471.77900\pm0.00054$					& 5 		& $-86 \pm 47 $	& -1.8 		& $-80 \pm 47$ 	& -1.7\\
20050414 	& $2453474.64398\pm0.00089$					& 7 		& $-84 \pm 77 $	& -1.1		& $-78 \pm 77$ 	& -1.0 \\
20060318		& $2453812.70871\pm0.00046$					& 7 		& $-14 \pm 40 $	& -0.4		& $-28 \pm 40$ 	& -0.7 \\
20070130		& $2454130.71844\pm0.00024$					& 465 	& $25 \pm  21$		& 1.2 		& $5\pm 21$ 		& 0.2\\
20080225		& $2454521.78377\pm0.00024$					& 738 	& $12 \pm 21 $		& 0.6			& $1\pm 21$ 		& 0.0\\
20080424		& $2454580.51523\pm0.00028$					&779 	& $13 \pm 24$ 		& 0.6			& $5 \pm 24$ 		& 0.2 \\
20080514		& $2454600.56983\pm0.00011$					& 793 	& $11 \pm 9$ 		& 1.2			& $3 \pm 9$ 		& 0.3 \\
20090315		& $2454905.68672\pm0.00032$					& 1006 	& $-1 \pm 28$		& -0.0 		& $13 \pm 28$ 		& 0.5\\
20090510		& $2454961.55291\pm0.00015$					& 1045 	& $-26 \pm 13$	 	& -2.0		& $-8 \pm 13$ 		& -0.6
\enddata
\tablenotetext{a} {Calculated from constant-period ephemeris Equation~\ref{ogle113eqn2}.}
\tablenotetext{b} {Calculated from linearly-changing-period ephemeris Equation~\ref{ogle113eqn4}.}
\tablenotetext{c} {Not refit; midtime reported by \citet{Konacki2004}.}
\label{table:ogle113ominusc}
\end{deluxetable}

\clearpage

\end{document}